\documentclass[pre,twocolumn,amsmath,showpacs]{revtex4}
\usepackage{graphicx}
\usepackage[dvips]{color}
\usepackage{bm}

\begin{document}

\title{Finite Width Model Sequence Comparison}

\author{Nicholas Chia}
\author{Ralf Bundschuh}
\email{bundschuh@mps.ohio-state.edu}

\affiliation{Department of Physics,
Ohio State University, 174 W. 18th St.,
Columbus, OH  43210, USA}

\date{\today}

\begin{abstract}
  Sequence comparison is a widely used computational technique in modern
  molecular biology. In spite of the frequent use of sequence comparisons the
  important problem of assigning statistical significance to a given degree of
  similarity is still outstanding. Analytical approaches to filling this gap
  usually make use of an approximation that neglects certain correlations in
  the disorder underlying the sequence comparison algorithm. Here, we
  use the longest common subsequence problem, a prototype sequence comparison
  problem, to analytically establish that this approximation does make a
  difference to certain sequence comparison statistics. In the course of 
  establishing this difference we develop a method that can systematically 
  deal with these disorder correlations.
\end{abstract}


\maketitle

\section{Introduction}

Sequence comparison gathers interest from a wide variety of fields
such as molecular biology, biophysics, mathematics, and computer
science. Methods of comparison concern computer scientists who use
string comparison for everything from file searches to image
processing \cite{Paterson,danc94a,danc94b,wate94}. Biological sequence
comparison provides details into the building blocks of life by
allowing the functional identification of newly found sequences
through similarity to already studied ones. Thus, it has become a
standard tool of modern molecular biology.

As with all pattern search algorithms, a crucial component for the
successful application of sequence comparisons is the ability to
discern the biologically meaningful from randomly occurring patterns. 
Thus, a thorough characterization of the strength of patterns within
\emph{random data} is mandatory to establishing a criterion for
discerning meaningful data \cite{karl90}.

The most commonly used sequence alignment algorithms are the closely
related Needleman-Wunsch~\cite{need70} and
Smith-Waterman~\cite{smit81} algorithms. There have been numerous
numerical and analytical studies that attempt to characterize the
behavior of these algorithms on random sequence
data~\cite{wate94a,wate94b,alts96, olse99,mott99,mott00,
sieg00,bund02,metz02}. However, there are
difficulties with both kinds of approaches to the problem of
characterizing sequence alignment algorithms statistically. The
numerical methods are by far too slow to be useful in an environment
where tens of thousands of searches are performed on a daily basis and
users expect their results on interactive time scales. The analytical
methods on the other hand, while in principle able to rapidly
characterize sequence alignment statistics, are only valid in small
regions of the vast parameter space of the sequence alignment
algorithms.

In addition to being restricted to a small region of parameter space, 
current analytical methods have another drawback: they rely
on an approximation to the actual alignment algorithm that ignores
some subtle correlations within the sequence disorder. Here, we want
to demonstrate that such correlations do matter and propose an 
analytical approach that can in principle deal with these correlations
for certain finite size variants of sequence alignment introduced in
section~\ref{sec:outline}.

We will concentrate on the simplest prototype of a sequence alignment
algorithm, namely the longest common subsequence (LCS) problem.  More
complicated models of sequence alignment can be adapted to the
methodology presented here in a straightforward manner. However, in
the interest of clarity and efficiency, we proceed with the simple
LCS problem in mind.  In
the longest common subsequence problem similarity between two randomly
chosen sequences over an alphabet of size $c$ is measured by the
length of the longest string that can be constructed from both
sequences solely by deleting letters. The central quantity
characterizing the statistics of the LCS problem is the expected
length of this longest common subsequence. Its fraction of the total
sequence length in the limit of infinitely long sequences is called
the Chv\'atal-Sankoff constant $a_c$.

Although the LCS problem is one of the simplest alignment algorithms,
the value of the Chv\'atal-Sankoff constant has been remarkably
elusive. So far, analytical stabs at $a_c$ have led to exact solutions
for very short lengths \cite{chva75} and proofs for upper and lower
bounds
\cite{Paterson,chva75,hirs78,chva83,deke83,danc94a,danc94b,alex94}.
Based on numerical results, there existed a long-standing conjecture
for the value of the Chv\'atal-Sankoff
constant. Recently~\cite{demo99,bund00}, this conjecture has been
proven to hold true for the approximation to the LCS problem that
precisely ignores the disorder correlations mentioned above.
Very careful and extensive numerical
treatments~\cite{danc94a,danc94b,demo99,demo00,bund01,dras01} 
have revealed that
the true Chv\'atal-Sankoff constant (including all disorder correlations)
deviates slightly from its value in the uncorrelated approximation.
This paper seeks to introduce a systematic way of understanding the
LCS problem with all disorder correlations included and to establish in an
analytically tractable environment that uncorrelated and correlated
disorder indeed lead to different results.

The format of this paper will be to summarize the LCS problem in
section~\ref{sec:review}. This section includes a general description of the
LCS problem and outlines a commonly used paradigm for solving for the LCS. In
addition, several conventions which are utilized throughout the paper are
defined here. In section~\ref{sec:outline} we introduce the finite width model
(FWM) method. In order to spare the reader possibly distracting mathematical
details we discuss only the overall ideas in the main text and reserve
appendix~\ref{sec:howto} for the more detailed discussion of the mathematical
methods employed in FWM. In section~\ref{sec:results} we give the results of
the FWM method for the correlated and uncorrelated LCS problem and discuss the
differences between these two problems that become obvious in the FWM
treatment.  Section~\ref{sec:conclusion} summarizes our findings.

\section{Review of the LCS problem}\label{sec:review}

\begin{figure}
\begin{center}
\includegraphics[width=0.9\columnwidth]{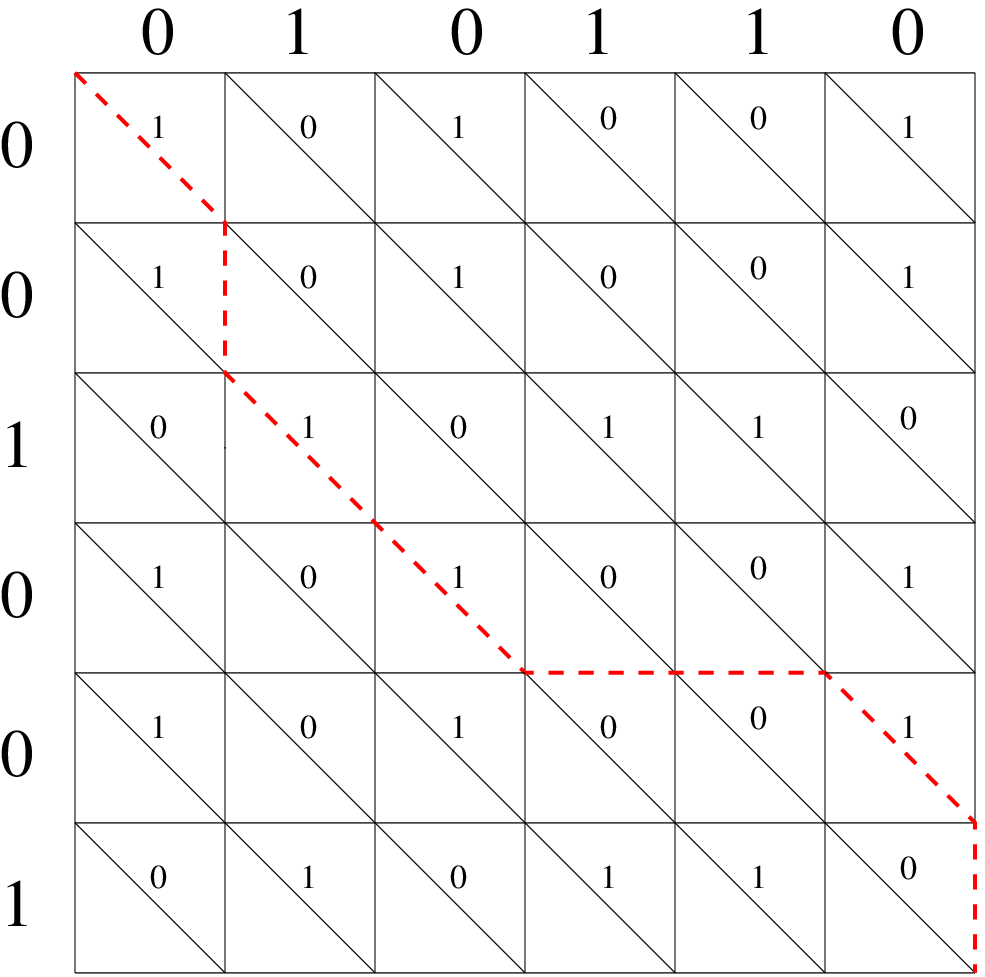}
\end{center}
\caption{\label{grid}Grid representation of the longest common subsequence
  (LCS) problem. The dashed line highlights a solution to the LCS 
  of the two binary sequences on the edges of the square. Notice that there
  exist multiple solutions to this problem.}
\end{figure}

The LCS of two sequences is the longest sequence that can be formed
solely by deletions in both sequences~\cite{chva75}. Best described by
example, the LCS of \verb+'DARLING'+ and \verb+'AIRLINE'+ is
\verb+'ARLIN'+, with a subsequence length of 5. Given two sequences of
length $M$ and $N$, $x_1 x_2...x_M$ and $y_1 y_2...y_N$, over an
alphabet of size c, their LCS can be computed in $O(MN)$ time. This
computation may be conveniently visualized with a rectangular grid
such as the one shown in Fig.~\ref{grid}. In this example, for the two
sequences, $x_1 x_2...x_6$ = \verb+'001001'+ and $y_1 y_2...y_6$ =
\verb+'010110'+, the LCS, \verb+'0100'+, has a length of 4. Within the
grid used to find the LCS, all horizontal and vertical bonds are
assigned a value of 0. Each diagonal bond is designated a value
depending on the associated letters in its row and column. Matching
letters earn their diagonal bonds a value of 1, while non-matching
letters result in an assignation of 0. Then, each directed path across
these various bonds from the first lattice point in the upper-left to
the last in the lower-right as drawn in Fig.~\ref{grid} corresponds to
a common subsequence of the two sequences. The only restriction is
that the path may never proceed against the order of the sequences. It
may only move rightward, downward, or right-downward in
Fig.~\ref{grid}. The length of a common subsequence corresponding to a
path is the sum of the bonds which comprise that path. Solving this
visual game for the length of the LCS requires that we find the path
of greatest value. This value will be the length of the LCS.

Recursively, we may define this problem by introducing the quantity 
$\ell (i,j)$ as the LCS of the substrings $x_1 x_2...x_i$ and 
$y_1 y_2 ... y_j$. Defining the LCS of two substrings in this way 
allows us to find the LCS leading to each of the lattice points in 
Fig.~\ref{grid}. This in turn breaks our path search down into 
the more manageable steps. 

\begin{eqnarray}\label{eq:recursion}
\ell (i,j) = \max \left\{ \begin{array}{l}
\ell (i-1,j) \\
\ell (i-1,j-1) + \eta (x,y) \\
\ell (i,j-1)
\end{array} \right\}
\end{eqnarray}
where
\begin{eqnarray}\label{eq:eta}
\eta (i,j) = \left\{ \begin{array}{ll}
1 & x_i = y_j \\
0 & \mbox{otherwise}
\end{array} \right.
\end{eqnarray}
and
\begin{equation}\label{eq:boundary}
\ell (0,j) = \ell (i,0) = 0.
\end{equation}

Of course, once we evaluate the final $k(M,N)$, we have solved 
for the length of the LCS. Ultimately, we wish to evaluate the 
central quantity that characterizes the LCS problem, the 
Chv\'atal-Sankoff constant $a_c$. It characterizes the ensemble 
of LCS's of pairs of randomly chosen sequences with $M=N$ 
independently identically distributed (iid) letters. If we denote 
averages over the ensemble by $\langle ... \rangle_N$, the 
Chv\'atal-Sankoff constant may be defined as
\begin{equation}
a_c \equiv
\underset{N \rightarrow \infty}{\lim} \frac{\langle LCS \rangle_N}{N} 
\end{equation}
This can be interpreted as the average growth rate of the LCS of 
two random sequences. 

As evident from Fig. 1 and the recursive equations 
(\ref{eq:recursion})-(\ref{eq:boundary}) the length of the LCS 
depends on the sequences only via the values of the $\eta$'s.
If we define the probabilities of 0 or 1 occurring in our random 
sequences as $p$ or $1-p$, respectively, where $0 \leq p \leq 1$, 
each individual $\eta$ carries a $2(1-p)p$ probability of being zero, 
and a $p^2 + (1-p)^2$ of being 1. However, the different $\eta (i,j)$
are not chosen independently according to those probabilities, but
are subject to subtle correlations.

It is very tempting to neglect these correlations in favor of
choosing $N^2$ iid variables $\widehat{\eta} (i,j)$ according to
\begin{eqnarray}
\widehat{\eta} (i,j) = \left\{ \begin{array}{lll}
1 & \mbox{with probability} & 1-q\\
0 & \mbox{with probability} & q
\end{array}
\right.
\label{eq:q}
\end{eqnarray}
where $q=2(1-p)p$, the probability of a bond value being 0. We will
call this the uncorrelated LCS problem and identify all quantities
calculated for this problem by an additional hat; specifically, we
will call $\widehat{a}_c$ the analog of the Chv\'atal-Sankoff constant
in the uncorrelated LCS problem. This approximation to the real LCS
problem has been used in various theoretical approaches to sequence
comparison statistics~\cite{mott99,demo99,bund02}. For the LCS problem
itself, only very careful numerical studies could show that the
Chv\'atal-Sankoff constant for the correlated and uncorrelated problem
are actually different \cite{danc94a,danc94b,demo99,demo00}.  However,
there are very real differences. The differences arise due to the fact
that the correlated and uncorrelated cases allow different sets of
possible bond or $\eta$ values. In the uncorrelated case all
combinations of bond values are allowed to exist. The grid in
Fig.~\ref{grid} makes it obvious that there are $NM$ bonds each with 2
possibilities. Therefore, there must exist $2^{NM}$ unique
configurations of bond values. Meager by comparison are the $2^{M + N
- 1}$ cases allowed by the two sequences of length M and N, with each
letter having the capacity to take on one of two values in the
correlated case. Notice the missing factor of 2 in the correlated case
arises due to the fact that one can alway replace all 0's with 1's in
order to get the same sequences of matches cutting our possible number
of bond value configurations in half. A more concrete realization of
the limited possibilities in the correlated case comes simply by
noticing that there are only two bond value configurations values any
row or column can take up in Fig.~\ref{grid}. Additionally, a column
or row with a value of 0 must be a mirror opposite of a column or row
with a value of 1. And so we can see that not only does the
uncorrelated case account for sets of bond values that cannot exist,
it cannot mimic the specific relationship between different rows and
columns of bond values. We will reveal these differences in a simple
approximation to the LCS problem that allows for closed analytical
solutions in the correlated and uncorrelated case.

\section{Finite Width LCS}\label{sec:outline}

\begin{figure}
\begin{center}
\includegraphics{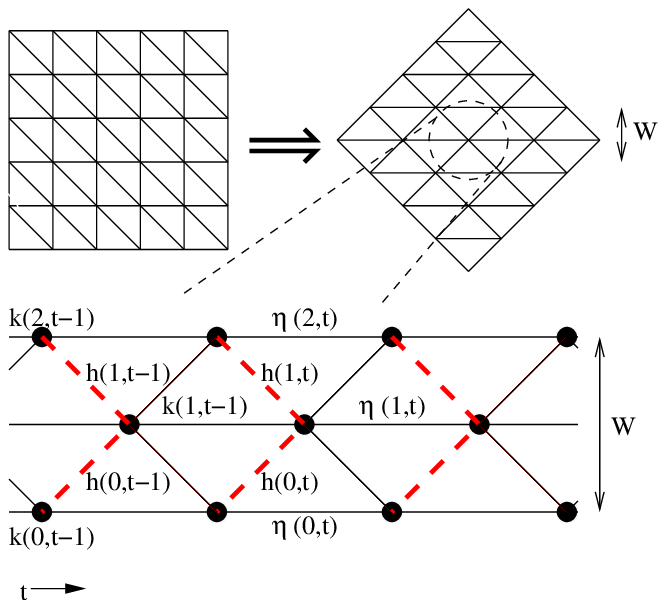}
\end{center}
\caption{\label{fwm}This picture shows our $45^o$ counter-clockwise 
rotation to achieve the orientation from which we will proceed. The 
blow up defines the lattice site values (k-values), the match values 
($\eta$-values), and the lattice site difference values (h-values). It 
also defines our time and width axes. The dashed lines connect the 
sites between which our h-values are measured.}
\end{figure}

The Finite Width Model (FWM) we will use in the following overlays the
grid presented in Fig.~\ref{grid} with the restriction in width
presented in Fig.~\ref{fwm}~\cite{bund00,bund01}. We will measure the
width $W$ of such a grid by the number of bands that make up the
lattice, i.e. $W=2$ in Fig.~\ref{fwm}.  Although, the grid used to
analyze the LCS must be truncated to a finite width W, our finite
strip extends to an infinite length. Thus, we can still define a width
dependent Chv\'atal-Sankoff constant
\begin{equation}
a_c (W) = \underset{N \rightarrow \infty}{\lim}
\frac{\langle \ell (N,N) \rangle_N}{N}
\end{equation}
In addition to the width $W$, the growth rate also depends on the
sequence composition. In our case of alphabet size $c=2$ this is
characterized by the probability $p$ to find a \verb+'0'+ on each site
within the sequences. While the method we will present below in
principle enables us to calculate any $a_{c}(W,p)$, we will, in the
following, concentrate on the simple example $a_{c}(W=2,p)$ shown in
Fig.~\ref{fwm}. Notice that
\begin{equation}
a_c = \underset{W \rightarrow \infty}{\lim}a_{c}(W,\frac{1}{2})
\end{equation}
from below, and thus $a_c(W,\frac{1}{2})$ produces a series of
lower bounds to the Chv\'atal-Sankoff constant.

On the finite width lattice shown in Fig.~\ref{fwm}, it is convenient
to redefine our quantities. Aside from the narrower scope under which
we investigate the LCS problem, all other properties of the grid
problem remain the same. Instead of referring to our lattice points by
the coordinates $i$ and $j$, we now utilize a time axis, $t$, which
points along the allowed, sequence forward, direction as well as a
coordinate axis, $x$, which lies perpendicular to the $t$-axis. 
In place of old $\ell$-values, in this new coordinate system we 
introduce $k$-values.
Keeping track of these $k$-values can be simplified to a new
set of recursive relationships at each time step. In the notation
defined by Fig.~\ref{fwm} for width $W=2$,
\begin{eqnarray}\label{eq:topline}
k(2,t+1) = \max \left\{ \begin{array}{l}
k(2,t) + \eta (2,t) \\
k(1,t)
\end{array} \right\}
\end{eqnarray}
\begin{eqnarray}\label{eq:midline}
k(1,t+1) = \max \left\{ \begin{array}{l}
k(2,t+1) \\
k(1,t) + \eta (1,t) \\
k(0,t+1)
\end{array} \right\}
\end{eqnarray}
\begin{eqnarray}\label{eq:botline}
k(0,t+1) = \max \left\{ \begin{array}{l}
k(1,t) \\
k(0,t) + \eta (0,t)
\end{array} \right\}
\end{eqnarray}
where the $\eta$'s take on values of either 1 or 0 in the same way
that we assigned values to our diagonal lines in Fig.~\ref{grid}.
Our set of recursive relationships gives us the longest path value
up to each lattice site. The length of the FWM LCS then becomes 
$k(1,N)$.

Related to our $k$-values by equations (\ref{eq:h1}) and 
(\ref{eq:h0}), we define the $h$-values in order to describe the 
relative values of our lattice sites within any given time frame.
Utilizing the diagrammed definitions of Fig.~\ref{fwm},
\begin{eqnarray}
h(1,t) = k(1,t)-k(2,t) \label{eq:h1} \\
h(0,t) = k(1,t)-k(0,t). \label{eq:h0}
\end{eqnarray}
The recursion relations (\ref{eq:topline}),(\ref{eq:midline})
and (\ref{eq:botline}), can be expressed entirely via this new 
quantity as
\begin{eqnarray}
h(1,t+1) = \max \left\{ \begin{array}{l}
0 \\
h(1,t) + \eta (1,t) - r(1,t) \\
h(1,t) + s(0,t) - r(1,t)
\end{array} \right\}
\label{eq:h1rec}
\end{eqnarray}
\begin{eqnarray}
h(0,t+1) = \max \left\{ \begin{array}{l}
0 \\
h(0,t) + \eta (0,t) - r(0,t) \\
h(0,t) + s(1,t) - r(0,t)
\end{array} \right\}
\label{eq:h0rec}
\end{eqnarray}
where
\begin{eqnarray}
r(1,t) & = & \max \left\{
\eta (2,t), h(1,t)
\right\}\label{eq:r1def}\\
r(0,t) & = & \max \left\{
\eta (0,t), h(0,t)
\right\}\label{eq:r0def}\\
s(1,t) & = & \max \left\{
0,\eta (2,t) - h(1,t)
\right\}\label{eq:s1def}\\
s(0,t) & = & \max \left\{
0, \eta (0,t) - h(0,t)
\right\}\label{eq:s0def}
\end{eqnarray}

Several properties conveniently arise from these definitions.  First,
notice that the $h$-values are independent of the absolute
$k$-values. Furthermore, it may be shown by inspection that the
$h$-values may only take on the values $(0,1)$. Inspecting
Fig.~\ref{fwm}, we note that each adjacent set of $k$-values share a
nodal $k$. This node attaches itself to the adjacent sites via a bond of
value 0 or 1. Since the nodal $k$ holds only a single value and the
single bonds leading to the adjacent sites can only change this value
by +1, the only $h$-values allowed then become 0 or 1. Having detached
the absolute $k$-values from the FWM-LCS problem entirely, we may
further detach the entire FWM from the grid in
Fig.~\ref{grid}. Originally we noted that our FWM-LCS on this grid
becomes $k(1,N)$. However, in order to calculate $a_c(W,p)$, the
length of the LCS problem has to be increased infinitely along the
time axis. Time now becomes an unbounded axis in the FWM. Since the
difference $|k(x,t)-k(y,t)|$ is bounded by $W$ for all $x$ and $y$,
the length of the LCS may be measured at any $x$. The growth rate may
then be expressed as the average $k$-values increase along any
coordinate, i.e.
\begin{equation}
a_c(W,p) = \langle k(x,t) - k(x,t-1) \rangle
\label{eq:grate}
\end{equation}
Notice that $k(x,t) - k(x,t-1) \epsilon \left\{ 0,1 \right\}$ and 
that as a result our newly defined $a_c(W,p)$ carries the condition
$0 \leq a_c(W,p) \leq 1$.

\begin{figure}
\begin{center}
\includegraphics[width=0.5\columnwidth]{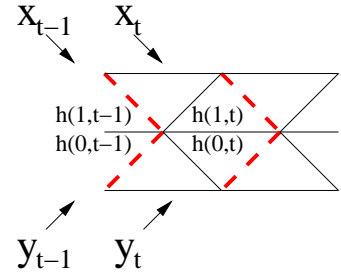}
\end{center}
\caption{\label{states} The diagram maps the influence of 
letters from the two sequences {..., $x_{t-1}$, $x_t$, ...} 
and {..., $y_{t-1}$, $y_t$, ...} on the new orientation. 
The letter information required for the evolution from time
$t-1$ to time $t$ is shown here. The dashed lines represent 
the chosen configuration for the $h$'s.} 
\end{figure}

The formulation given by equations (\ref{eq:h1rec}),
(\ref{eq:h0rec}) make it clear that $h(1,t)$ and $h(0,t)$ can be
calculated if $h(1,t-1), h(0,t-1), \eta (2,t-1), \eta (1,t-1)$,
and $\eta (0,t-1)$ are known. This allows us to write the time
evolution as a Markov model. In order to do so, the information
required for the time evolution at each time step must be 
included in the states. For our uncorrelated states, where the
$\eta$'s occur randomly, only $h(1,t-1)$ and $h(0,t-1)$ are
required to determine the probable time evolution. Therefore,
the uncorrelated states simply read ($h(1,t-1)$, $h(0,t-1)$). 
The probabilities for the time evolution into the state
($h(1,t)$, $h(0,t)$) may then be calculated based on the 
$\eta$-value probabilities given in equation (\ref{eq:eta}). 
However, in the correlated case, the $\eta$'s are not randomly
chosen. Instead, the letters in each sequence are according
to the probability $p$ of a 0 occurring at a single site within
the sequences. Some letters affect $\eta$'s across multiple 
time steps as shown by Fig.~\ref{states}. In order to
calculate the time evolution to time $t$, $\eta (2,t-1), 
\eta (1,t-1)$, and $\eta (0,t-1)$ must be known. These $\eta$'s
depend on the subsequences $x_{t-1},x_t$ and $y_{t-1},y_t$. Once
these four letters are known, the state at time $t$ may be
determined. Since these letters cast an influence across
multiple time steps, their information must be retained in order
to accurately forecast the upcoming possibilities for the 
$\eta$'s and our states. Redefining our states as
($h(1,t),h(0,t),x_t,y_t$) preserves the necessary information.
The remaining information for calculating the $\eta$'s
needed for the time evolution, mainly $x_{t+1}$ and $y_{t+1}$,
arise according to the letter probabilities. These probabilities
contribute to the probable time evolution into the next $t+1$ 
state ($h(1,t+1),h(0,t+1),x_{t+1},y_{t+1}$). It can be shown
that correlated states must always contain $W$ $h$-values and $W$
letter values, and that uncorrelated states must always contain
$W$ $h$-values.

\begin{figure}
\begin{center}
\includegraphics{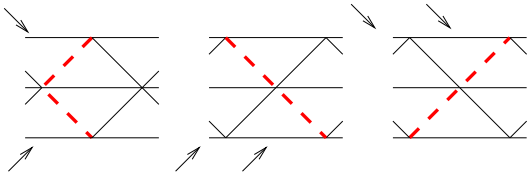}
\end{center}
\caption{\label{shapes}The dashed lines represent the various 
configurations by which the $h$-values can be defined. Note 
that each of the various sets of $h$-values implies a different 
definition of the state, and thus requires a different set of 
letters. The arrows represent the letters which are required in 
each different configuration.}
\end{figure}

Though the number of elements in a state depends only on the 
width, there exist alternative means of writing our states. We
are free to choose whatever configuration of continuous lines
to define our $h$-values across. Fig.~\ref{shapes} show the
other possible configurations in width 2 FWM. Naturally, the
letter effects differ for each shape, and the proper letters
for each configuration are also illustrated. The various states
we may form all contain the same number of $h$ and letter 
values and obey the same principles. Only the specified set of 
$h$ and letter values differ.

Whatever state we choose to define, the Markov process \ describing
FWM-LCS is characterized by a transfer matrix $\hat{T}$. This matrix
describes the transitions from a state in one time to a state in it's
immediate future.  It is a representation of the dynamics given by
Eqs.~(\ref{eq:h1rec})-(\ref{eq:s0def}).  We leave the mechanics of
obtaining this transfer matrix to Appendix \ref{sec:howto} and focus
here on the results.

Once we have found the transfer matrix we may solve for the 
vector ${\vec s}$ describing the steady state by solving the 
linear system of eigenvalue equations
\begin{equation}
{\hat T} \cdot {\vec s} = {\vec s}
\label{eq:eigen}
\end{equation}
subject to the normalization condition 
${\vec 1} \cdot {\vec s} = 1$  where 
${\vec 1} = (1, 1, 1, 1, ...)$. Note that the size of these vectors 
depends on the number of states needed to describe the problem. More 
specifically, 16 elements are needed for the correlated width 2 FWM 
while the uncorrelated width 2 case only requires 4 elements. This 
steady state vector must contain the probabilities to observe 
every single state in the random ensemble. Note that the directness 
of this technique allows for it's ready adaptation to more complex 
sequence comparison algorithms along the lines of Ref.~\cite{bund02}.
However, this generally requires a significantly
larger number of states thus incurring a greater computational 
cost. 

In order to describe the growth rate, we utilize a growth 
matrix $\hat{G}$ to mark the transitions which result in growth 
along some chosen coordinate. The process by which we construct
the growth matrix bears great similarity to the process by
which we construct the transfer matrix. In fact, the growth
matrix only omits those elements of the transfer matrix that 
do not contribute to the growth along a chosen coordinate.
Further detail regarding the construction of the growth matrix
has been left for Appendix \ref{sec:howto}.

The growth matrix allows us to define the growth vector, ${\vec g}$
\begin{equation}
{\vec g} = {\vec 1} \cdot {\hat G}
\label{eq:growth}
\end{equation}
This growth vector describes the probable growth from each 
of the states. Coupled with the steady state, which provides 
us with the likelihood of each state, this allows us to solve 
for $a_c(W,p)$ directly as
\begin{equation}
a_c(W,p) = {\vec g} \cdot {\vec s}
\label{eq:LCS}
\end{equation}
since the probability of growth from $t \rightarrow t+1$, as
described by the growth matrix, is independent of the probability
to be in a certain state at time $t$.

Before we discuss the results of this approach, we would like to point
out that this technique is not limited to the calculation of the
growth rate $a_c(W,p)$. Since the dynamics of the scores is a Markov
process, any quantity can be calculated once the transfer matrix $\hat
T$ and the steady state vector $\vec s$ are known. E.g., any
equal-time correlation function of interest can be obtained directly
from the steady-state vector $\vec s$ simply by summing over the
degrees of freedom that are not to be included in the correlation
function while a time-correlation function like $\langle
i(t)|j(t')\rangle$ (the probability to be in state $i$ at time $t$
given that the system was in state $j$ at time $t'$) is simply given
by $\langle i(t)|j(t')\rangle=\vec i\cdot\hat T^{t-t'}\vec j$ where
$\vec i$ and $\vec j$ are vectors, all entries of which are zero
except for a one in the row for state $i$ or $j$, respectively.

Solving the correlated width $W=2$ FWM-LCS problem utilizing the 
process given by equations (\ref{eq:eigen})-(\ref{eq:LCS}), we 
arrive at the equation
\begin{equation}
a_{2}(W=2,p) = 
\frac{3 - 5 p + 5 p^2}{3 - p - 3 p^2 + 8 p^3 - 4 p^4}
\label{eq:LCSc}
\end{equation}
where $p$ represents the probability of the first letter 
occurring. The same methodology may be applied to the 
uncorrelated case, where we describe the transition 
probabilities using the bond probability q defined by equation 
(\ref{eq:q}).
\begin{equation}
\widehat{a}_{2}(2,q) = 
\frac{5 - 7 q + 2 q^2}{5 - 5 q + q^2}
\label{eq:LCSu1}
\end{equation}
Notice that the specifics of the state that we choose does not 
impact the result in any way. Nor does the choice we make with 
respect to measuring the growth. Any combination of choices result 
in equations (\ref{eq:LCSc}) and (\ref{eq:LCSu1}) for the correlated
and uncorrelated cases respectively. We explicitly verified this 
independence in the choice of configurations and definitions of 
the growth. These independent results serve as a powerful check 
for the correctness of the algebraic manipulations. Substituting 
$q = 2(1-p)p$ into equation (\ref{eq:LCSu2}), the probability of 
getting two different letters, or a bond value of 0, gives a
an equation expressed in the same quantities as the correlated
Chv\'atal-Sankoff constant given by equation (\ref{eq:LCSc}), 
mainly
\begin{equation}
\widehat{a}_{2}(2,p) = 
\frac{5 - 14 p + 22 p^2 - 16 p^3 + 8 p^4}
{5 - 10 p +14 p^2 - 8 p^3 + 4 p^4}\label{eq:LCSu2}
\end{equation}

\begin{figure}[b]
\begin{center}
\includegraphics[width=0.9\columnwidth]{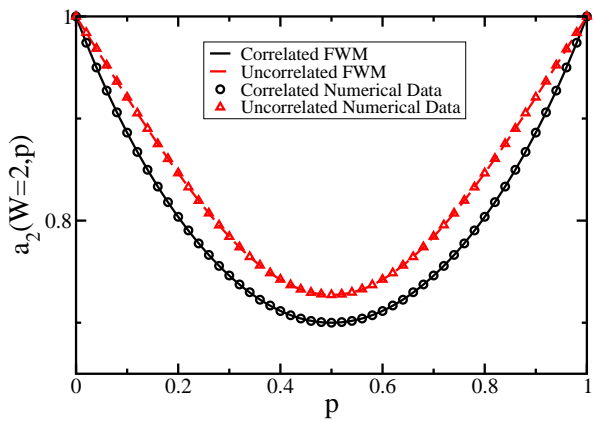}
\end{center}
\caption{\label{plot}Analytical and numerical data provided by
FWM. This plot shows further evidence verifying the correctness of 
FWM. Numerical modeling obtained by passing many random sequences 
through an FWM evaluation produces the data points represented. 
The analytical FWM model matches the numerical data with high 
precision for both the correlated and uncorrelated cases at width
$W=2$. The error for the numerical data presented is smaller
than the symbol size.}
\end{figure}

\section{Results}\label{sec:results}

Now, we will apply our method to various small width cases and discuss
the implications of the results for the longest common subsequence
problem.  First, we check our computations, and plot the results
Eqs.~(\ref{eq:LCSc}) and~(\ref{eq:LCSu2}) alongside numerical data
obtained by random sampling in Fig.~\ref{plot}. The numerical data
obtained by choosing $10,000$ pairs of random sequences of length
$10,000$, calculating their width $2$ LCS and averaging shows no
discernable deviation from the analytical results over the whole range
of the parameter $p$.  Already in this plot for $W=2$, the differences
between the correlated and uncorrelated cases are apparent.
Coinciding only for $p=0$ and $p=1$ where growth is certain in every
step, the two cases differ at all other points.

\begin{table*}
\begin{ruledtabular}
\begin{tabular}{lll}
width $W$ & correlated case & uncorrelated case \\
\hline
0 & $\frac{1}{2}$ (0.5) & $\frac{1}{2}$ (0.5) \\
1 & $\frac{2}{3}\ (0.\bar{6})$ & $\frac{2}{3}\ (0.\bar{6})$ \\
2 & $\frac{7}{10}$ (0.7) & $\frac{8}{11}\ (0.\bar{72})$\\
3 & $\frac{1592}{2201}$ (0.723307587) & $\frac{34}{45}\ (0.7\bar{5})$\\
4 & $\frac{3900482569}{5288762638}$ (0.737503808) & $\frac{152}{197}$ (0.771573604)\\
5 & $\frac{1016932681760084189805278879341973703014985562}{1359136362951380586870384955918322158719785917}$ (0.748219759) & $\frac{706}{903}$ (0.781838317)\\
$\infty$ & (0.8126) & (0.828427125)
\end{tabular}
\end{ruledtabular}
\caption{\label{tab:growth} Finite-width Chv\'atal-Sankoff constant
$a_2(W,\frac{1}{2})$ for correlated and
$\widehat{a}_2(W,\frac{1}{2})$ for uncorrelated disorder.
As a reference the value of the Chv\'atal-Sankoff constant for 
infinite width is given. In the correlated case it is only known 
numerically~\protect\cite{bund01}; in the uncorrelated case it is given by 
$2/(\sqrt{2}+1)$~\protect\cite{demo99,bund00}.}
\end{table*}

Then, we look at the width dependence of the growth rates at the symmetric
point $p=1/2$. They are summarized in Table \ref{tab:growth}. The results
again verify the difference between the correlated and uncorrelated cases with
the growth rate in the uncorrelated case being systematically higher than in
the correlated case. They also highlight two rather interesting exceptions.
The first occurs for the case $W=0$ in which correlations play no role and
indeed have no meaning. Assigning a random bond value (uncorrelated) or two
random letters (correlated) lead to the same effect. Thus, as expected, the
correlated and uncorrelated cases plot identically for $W=0$. The second
exception, occurring for $W=1$, narrows the scope of equality to three values
of $p$, namely $0,\frac{1}{2},$ and $1$.  In these cases, the equality arises
from the exactly similar bond values being produced from each case. In all
other respects, the correlated and uncorrelated versions of $W=1$ differ.

Viewing the solutions of Table \ref{tab:growth} also shows that the growth
rate $a_c$ increases with width $W$. This agrees with perfectly with the
expectations of the FWM. As width increases so do the possibilities for
growth. In fact, in the limit $W \rightarrow \infty$ we recover the
Chv\'atal-Sankoff constant - the infinite width growth rate. Along the way,
these finite width values of $a_c$ provide lower bounds to the
Chv\'atal-Sankoff constant. In this way, FWM conveniently provides a method for
gathering systematic solutions for obtaining lower bounds to the
Chv\'atal-Sankoff constant. The values given for $a_c$ in this table may be
read as a series of ever increasing analytically solved lower bounds. In this
systematics, FWM displays one of it's advantages over conventional methods.
However, the power and exactness of these solutions exacts a computational
cost that grows as $2^{12W}$.

\begin{figure}[b]
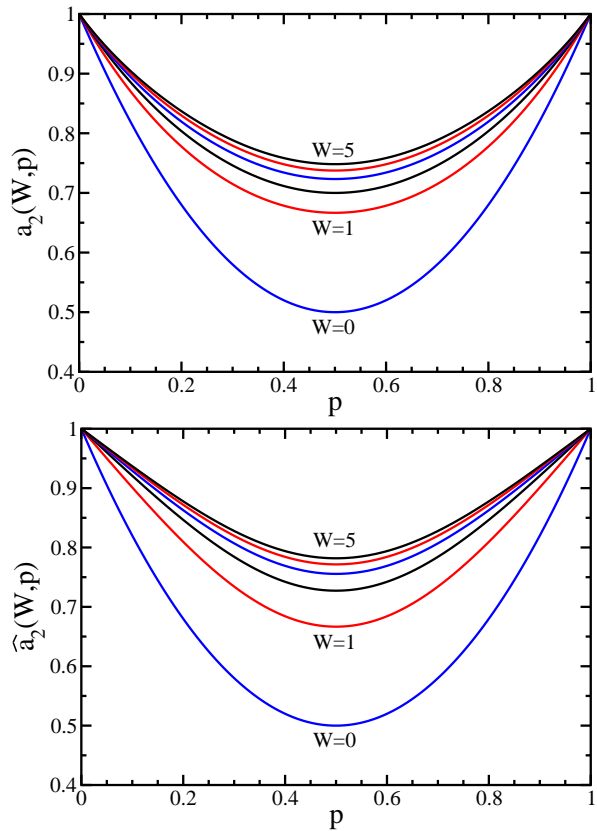

\includegraphics[width=0.9\columnwidth]{6a.eps}
\includegraphics[width=0.9\columnwidth]{6b.eps}
\caption{\label{allwidths}Finite width growth rates as a function 
of the letter probability $p$ for different widths.}
\end{figure}

\begin{figure}
\includegraphics[width=0.9\columnwidth]{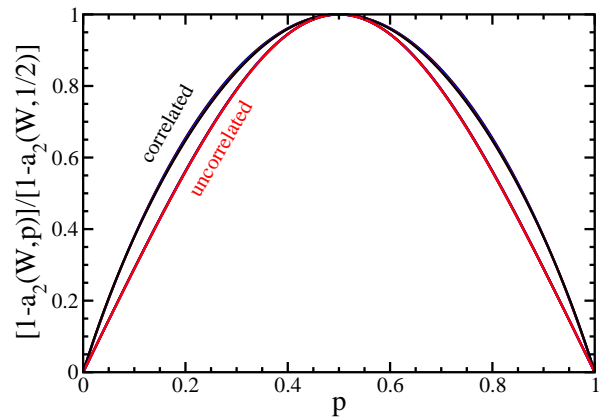}
\caption{\label{rescale}Rescaled growth rate for $W=2,3,4,5$ in the 
correlated and uncorrelated case as well as the $W=\infty$ result 
for the uncorrelated case. All results for the correlated case and 
all results of the uncorrelated case are virtually indistinguishable 
from each other while the correlated growth rate clearly follows a 
pattern that is distinctly different from the uncorrelated growth 
rate.}
\end{figure}

Next, we consider the dependence of the growth rates on the letter
probability $p$. Fig.~\ref{allwidths} shows the full analytical
solutions for various widths plotted as a function of $p$. These
graphs verify the trend noted from the discussion of Table
\ref{tab:growth}. However, they allow another interesting observation:
while the difference in values in the correlated and uncorrelated
cases may be immediately perceived, the shape of each of the curves
appears to not depend on $W$. With increasing $W$ the curve simply
appears to come closer and closer to one. In order to verify this, we
rescale the difference $1-a_c(W,p)$ of the growth rate from one by its
value $1-a_c(W,1/2)$ at $p=1/2$. As shown in Fig.~\ref{rescale} these
rescaled curves are indeed indistinguishable for $W = 2,3,4,$ and
$5$. They clearly fall into two distinct classes, namely a curve for
the correlated case and a curve for the uncorrelated case. For the
uncorrelated case, where the result
$\widehat{a}_c(W=\infty,p)=2/[(p^2+(1-p)^2)^{-1/2}+1]$ for infinite
width is known~\cite{demo99,bund00}, Fig.~\ref{rescale} also shows
perfect agreement between the finite $W$ and the infinite $W$
results. Thus, at least in the uncorrelated case there are no
noticeable finite size effects in the scaling function even for widths
as small as $W=2$. Assuming the absence of finite size effects even
for small widths also holds true for the correlated case for which we
cannot independently verify this assumption, the results shown in
Fig.~\ref{rescale} support two important conclusions: (i) the
correlated and uncorrelated systems truly and systematically differ
for all widths and thus also in the limit $W\to\infty$, and (ii) these
curve shapes can be understood as universal properties of the
correlated and uncorrelated FWM-LCS system independent of the width
$W$. It implies that, given the value of $a_c$ for any $p \not= 0$ or
$1$ one may plot $a_c$ for all values of p. In other words, a single
data point suffices to define a finite width system whether it be
correlated or uncorrelated.

The differences between the correlated and uncorrelated case, 
highlighted by Fig.~\ref{rescale}, result from the subtle 
restrictions that correlations place on bond values. As an example, 
for width 2 FWM, three bond values contribute to a single transition.
Thus $2^3 = 8$ unique sets of bond values exist. Uncorrelated bond
values allow for any of these $8$ possibilities at any given time.
However, because the letters effect correlated bonds in multiple
time steps, each correlated state has only $4$ allowed transitions. 
In fact, in any width the FWM provides a maximum of $4$ allowed 
transitions for all correlated states. The reasons for this are
elucidated in Appendix \ref{sec:howto}. In addition to the number of 
possibilities lacking in the correlated case, the allowed transitions
create subtle relationships creating patterns of growth that differ
significantly from the uncorrelated case. These differences account
for the systematic separation viewed in Fig.~\ref{rescale}.

\section{Conclusion}\label{sec:conclusion}

We conclude that within the FWM method differences between the
correlated and uncorrelated LCS problem can be established analytically.
The dependence of the finite width growth rate on the letter probability $p$
follows a scaling law already for the relatively small widths which are
analytically accessible. These scaling laws are distinctively different
for the correlated and uncorrelated case within FWM thereby providing an 
analytical argument that the differences between the correlated and 
uncorrelated case explicitly revealed for small finite widths here may 
persist in the limit of infinite widths. This is the first piece of 
analytical 
evidence that hints at the distinctness of the Chv\'atal-Sankoff constants 
in the correlated and uncorrelated cases. However, though there exists an
analytical 
solution for the infinite width uncorrelated case, it should be noted that
no such solution for the infinite width correlated case is available. Thus
this evidence has only been analytically verified for widths up to 5 for 
correlated finite width systems, and the pattern suggested by this data 
set may yet be the result of some finite width effect. 
Nonetheless, the FWM method in itself provides a systematic means to
deal with these correlations that can be generalized from the LCS to other
sequence comparison problems.

\begin{appendix}
\section{Obtaining the Transfer and Growth Matrices}\label{sec:howto}

\begin{figure}
\begin{center}
\includegraphics{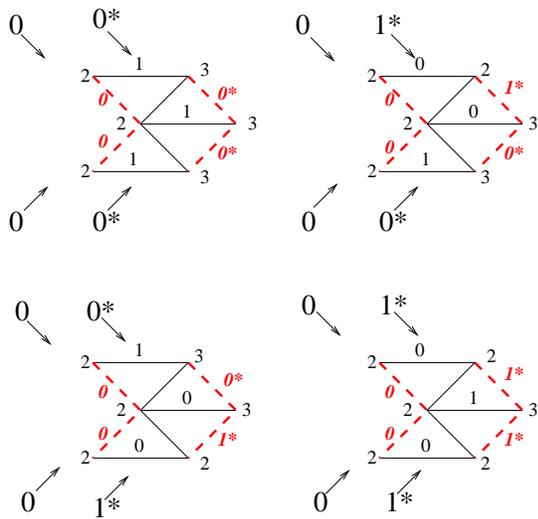}
\end{center}
\caption{\label{example}Above the four possible futures or 
transitions available to the state (0,0,0,0) are obtained 
diagrammatically. These transitions, reading from the
upper left, are (0,0,0,0), (1,0,1,0), (0,1,0,1) and 
(1,1,1,1). Note that the states are organized with h values 
first, then letters both written in from the top to the bottom 
in this diagram. In order to help clarify the origin of these 
four sets of numbers, the quantities relevant to the new 
states have been starred.}
\end{figure}

Our transfer matrix, as discussed in the main text, describes 
transitions from one state into the next. It allows us to determine 
the probable fraction of time spent in any state, i.e. the steady 
state, and coupled with the growth matrix it allows us to calculated 
the growth rate. Obtaining the matrix elements involves finding all
transition probabilities and placing them into our matrix. To begin, 
one simply takes a state and writes all possible transitions out of 
this state. When one has done this for all possible states, then the 
transfer matrix is complete. As an example we have calculated the 
first column of the transfer matrix in the correlated case $W=2$.

Starting with the first column, which represents our 
(0, 0, 0, 0) state, we note that there exist only four possible
futures. Once we choose the two remaining letters as (0, 0), (1, 0), 
(0, 1) or (1, 1) the differences $h$ are completely determined. 
Fig.~\ref{example}, shows the determination of the state 
transitions that result from these four sets of letters. These four 
transition then become the matrix elements of the first column. The 
probability weighing each transition is determined by the new set 
of letters that bring about the new state, or the starred letters 
in Fig.~\ref{example}. In the order listed above, the states 
they bring about are weighed by the probabilities $p^2$, $(1-p)p$, 
$p(1-p)$, and $(1-p)^2$.

In order to formulate a growth matrix, we pick the line defining 
the growth, and delete the elements of the transfer matrix which do 
not contribute to the growth on this line. In this example we have 
chosen to measure our growth along the bottom line. As an example, 
in Fig.~\ref{example}, the two top diagrams contribute to growth 
because the lattice value along the bottom line grows in both these 
cases. However, for the bottom pair, the lower lattice value 
remains static, thus their contributions are missing from the 
growth matrix shown below.

Repeating this for each possible starting state leads to the
following matrix representations where the states are ordered from
least to greatest in binary (0000, 0001, 0010, 0011,...).

\begin{widetext}
\def\omp{(1\!\!-\!\!p)}
\begin{tiny}
$$ \mbox{\large$\displaystyle{\hat T}=$}\left(
\begin{array}{cccccccccccccccc}
p^2 & 0 & 0 & 0 & 0 & 0 & 0 & 0 & 0 & 0 & 0 & 0 & 0 & 0 & 0 & 0 \\[1mm]
0 & \omp p & \omp p & 0 & 0 & \omp p & 0 & \omp p &\omp p & \omp p & 0 & 0 & \omp p & \omp p & \omp p & \omp p \\[1mm]
0 & \omp p & \omp p & 0 & \omp p & 0 & \omp p & 0 & 0 & 0 & \omp p & \omp p & \omp p & \omp p & \omp p & \omp p \\[1mm]
0 & 0 & 0 & \omp^2 & 0 & 0 & 0 & 0 &  0 & 0 & 0 & 0 & 0 & 0 & 0 & 0 \\[1mm]
0 & 0 & p^2 & 0 & p^2 & 0 & p^2 & 0 & 0 & 0 & 0 & 0 & 0 & 0 & 0 & 0 \\[1mm]
\omp p & 0 & 0 & 0 & \omp p & 0 & \omp p & 0 &  0 & 0 & 0 & 0 & 0 & 0 & 0 & 0 \\[1mm]
0 & 0 & 0 & \omp p & 0 & \omp p & 0 & \omp p &  0 & 0 & 0 & 0 & 0 & 0 & 0 & 0 \\[1mm]
0 & \omp^2 & 0 & 0 & 0 & \omp^2 & 0 & \omp^2 & 0 & 0 & 0 & 0 & 0 & 0 & 0 & 0 \\[1mm]
0 & p^2 & 0 & 0 & 0 & 0 & 0 & 0 & p^2 & p^2 & 0 & 0 & 0 & 0 & 0 & 0 \\[1mm]
0 & 0 & 0 & \omp p & 0 & 0 & 0 & 0 & 0 & 0 & \omp p & \omp p & 0 & 0 & 0 & 0 \\[1mm]
\omp p & 0 & 0 & 0 & 0 & 0 & 0 & 0 & \omp p & \omp p & 0 & 0 & 0 & 0 & 0 & 0 \\[1mm]
0 & 0 & \omp^2 & 0 & 0 & 0 & 0 & 0 & 0 & 0 & \omp^2 & \omp^2 & 0 & 0 & 0 & 0 \\[1mm]
0 & 0 & 0 & p^2 & 0 & p^2 & 0 & p^2 & 0 & 0 & p^2 & p^2 & p^2 & p^2 & p^2 & p^2 \\[1mm]
0 & 0 & 0 & 0 & 0 & 0 & 0 & 0 &  0 & 0 & 0 & 0 & 0 & 0 & 0 & 0 \\[1mm]
0 & 0 & 0 & 0 & 0 & 0 & 0 & 0 &  0 & 0 & 0 & 0 & 0 & 0 & 0 & 0 \\[1mm]
\omp^2 & 0 & 0 & 0 & \omp^2 & 0 & \omp^2 & 0 &  \omp^2 & \omp^2 & 0 & 0 & \omp^2 & \omp^2 & \omp^2 & \omp^2
\end{array}  \right)$$ \end{tiny}

\def\omp{(1\!\!-\!\!p)}
\begin{tiny}
$$ \mbox{\large$\displaystyle{\hat G} =$} \left(
\begin{array}{cccccccccccccccc}
p^2 & 0 & 0 & 0 & 0 & 0 & 0 & 0 & 0 & 0 & 0 & 0 & 0 & 0 & 0 & 0 \\[1mm]
0 & 0 & \omp p & 0 & 0 & \omp p & 0 & \omp p & 0 & 0 & 0 & 0 & \omp p & \omp p & \omp p & \omp p \\[1mm]
0 & \omp p & 0 & 0 & \omp p & 0 & \omp p & 0 & 0 & 0 & 0 & 0 & \omp p & \omp p & \omp p & \omp p \\[1mm]
0 & 0 & 0 & \omp^2 & 0 & 0 & 0 & 0 & 0 & 0 & 0 & 0 & 0 & 0 & 0 & 0 \\[1mm]
0 & 0 & 0 & 0 & p^2 & 0 & p^2 & 0 & 0 & 0 & 0 & 0 & 0 & 0 & 0 & 0 \\[1mm]
0 & 0 & 0 & 0 & \omp p & 0 & \omp p & 0 & 0 & 0 & 0 & 0 & 0 & 0 & 0 & 0 \\[1mm]
0 & 0 & 0 & 0 & 0 & \omp p & 0 & \omp p & 0 & 0 & 0 & 0 & 0 & 0 & 0 & 0 \\[1mm]
0 & 0 & 0 & 0 & 0 & \omp^2 & 0 & \omp^2 & 0 & 0 & 0 & 0 & 0 & 0 & 0 & 0 \\[1mm]
0 & p^2 & 0 & 0 & 0 & 0 & 0 & 0 & p^2 & p^2 & 0 & 0 & 0 & 0 & 0 & 0 \\[1mm]
0 & 0 & 0 & \omp p & 0 & 0 & 0 & 0 & 0 & 0 & \omp p & \omp p & 0 & 0 & 0 & 0 \\[1mm]
\omp p & 0 & 0 & 0 & 0 & 0 & 0 & 0 & \omp p & \omp p & 0 & 0 & 0 & 0 & 0 & 0 \\[1mm]
0 & 0 & \omp^2 & 0 & 0 & 0 & 0 & 0 & 0 & 0 & \omp^2 & \omp^2 & 0 & 0 & 0 & 0 \\[1mm]
0 & 0 & 0 & 0 & 0 & p^2 & 0 & p^2 & 0 & 0 & 0 & 0 & p^2 & p^2 & p^2 & p^2 \\[1mm]
0 & 0 & 0 & 0 & 0 & 0 & 0 & 0 & 0 & 0 & 0 & 0 & 0 & 0 & 0 & 0 \\[1mm]
0 & 0 & 0 & 0 & 0 & 0 & 0 & 0 & 0 & 0 & 0 & 0 & 0 & 0 & 0 & 0 \\[1mm]
0 & 0 & 0 & 0 & \omp^2 & 0 & \omp^2 & 0 & 0 & 0 & 0 & 0 & \omp^2 & \omp^2 & \omp^2 & \omp^2
\end{array} \right)$$ \end{tiny}

\end{widetext}

\end{appendix}

\end{document}